\documentclass[aps,prl,twocolumn,superscriptaddress,floatfix]{revtex4}

\usepackage{graphicx}
\usepackage[latin1]{inputenc}
\usepackage{amsmath}
\usepackage{amssymb}
\usepackage{color}

\newcommand{\nc}{\newcommand}
\nc{\bra}[1]{\langle #1|}
\nc{\ket}[1]{|#1\rangle}
\nc{\braket}[1]{\left\langle #1 \right\rangle}
\nc{\equ}[1]{\begin{eqnarray*}#1\end{eqnarray*}}
\nc{\equn}[1]{\begin{eqnarray}#1\end{eqnarray}}
\nc{\dagg}{^{\dagger}}
\nc{\conj}{^{*}}
\nc{\dx}[1]{\, \mathrm{d} {#1} \,}
\nc{\Dx}[1]{\mathcal{D} {#1} \,}
\nc{\la}{\langle}
\nc{\ra}{\rangle}
\nc{\tr}{\text{tr}}
\nc{\Tr}{\text{Tr} \,}
\nc{\e}{\text{e}}
\nc{\Id}{\mathbb{1}}
\nc{\eps}{\varepsilon}
\nc{\der}[2]{\frac{\mathrm{d} {#1}}{\mathrm{d} {#2}}}
\nc{\pder}[2]{\frac{\partial {#1}}{\partial {#2}}}
\nc{\bigO}{\mathcal{O}}
\nc{\half}{\frac{1}{2}}
\nc{\Eq}[1]{Eq.~(\ref{#1})}
\nc{\eq}[1]{Eq.~(\ref{#1})}
\nc{\chap}[1]{Chapter \ref{#1}}
\nc{\Sect}[1]{Section \ref{#1}}
\nc{\sect}[1]{section \ref{#1}}
\nc{\fig}[1]{Fig.~\ref{#1}}
\nc{\Fig}[1]{Fig.~ref{#1}}
\nc{\tabl}[1]{Table \ref{#1}}
\nc{\app}[1]{Appendix \ref{#1}}
\nc{\eg}{\emph{e.g.}}
\nc{\ie}{\emph{i.e.}}
\nc{\etal}{\emph{et al.}}
\nc{\asp}[1]{\textcolor{red}{[{\bf SP}: #1]}}
\nc{\alg}[1]{#1}

\begin{document}

\title{Quantum correlations in predictive processes}

\author{Arne L. Grimsmo}\email{arne.grimsmo@ntnu.no}
\affiliation{Department of Physics, The Norwegian University of Science and Technology, N-7491 Trondheim, Norway}
\affiliation{Department of Physics, University of Auckland, Private Bag 92019, Auckland, New Zealand}

\date{\today}

\begin{abstract}
  We consider the role of quantum correlations in the efficient use of information by a predictive quantum system, generalizing a recently proposed classical measure of non-predictive information to the quantum regime. We show that, as a quantum system changes state, the non-predictive information held by another correlated quantum system is exactly equal to the extractable work that is lost from the second system. We use quantum discord to quantify the quantum contribution, and demonstrate the possibility of improved thermodynamic efficiency due to a negative ``quantum part'' of the lost work. We also give a thermodynamic interpretation to quantum discord, as the reduction in extractable work under an optimal classical approximation of \alg{a quantum memory}.
\end{abstract}

\pacs{}

\maketitle

A system which by interacting with its environment tries to predict the future of some part of its surroundings is essentially involved in the process of learning. In a realistic situation, the changing environment must be modeled as a stochastic process. From an information theory point of view, it may be thought of as a source emitting a random message, $x$, with probability $p_x(t)$. The system changes its state in response to this message and can be thought to perform an implicit computation of the environment's variables \cite{Still12}; the states of the system and environment become correlated. As the evolution of the environment exhibits temporal correlations, the system becomes correlated with the past of its surroundings and also, implicitly, with their future. From a practical perspective, the most useful model is one that posseses predictive power without being unnecessarily complex---as much retained information as possible should be useful for predicting the future of the surroundings \cite{Still12,Still10}. In a recent publication, Still \emph{et al.} \cite{Still12} showed that an information theoretic measure of the inefficiency of the predictive process (the non-predictive information) is, under certain conditions, equal to thermodynamic inefficiency, \ie, the dissipated work during the evolution. The result shows a deep connection between effective use of information and efficient thermodynamic operation.

Their discussion was limited to the case where both system and environment are classical. In this paper we make an extension to quantum parties. As motivation we note, first, that real environments are quantum, so quantum effects must come into play for microscopic systems at some stage. Second, we might ask whether there can be a non-classical advantage in the effective use of information, and thus in thermodynamic efficiency.

Considering systems that are quantum, classical states must be replaced by quantum states---for uncertain $x$, by mixed states $\rho(t) = \sum_x p_x(t) \ket{\psi_x(t)}\bra{\psi_x(t)}$; importantly, quantum states are not distinguishable in general. Furthermore, quantum systems can be correlated in ways that are inaccessible to a classical stochastic processes, \emph{e.g.}, through non-local entanglement \cite{Bell64}. It is natural then to ask what type of correlations might constitute a resource for efficient operation in the quantum case.

We show that a relation between lost work and predictive inefficiency holds also for quantum systems, in analogy with the classical case. 
Since the shared information may now involve quantum correlations, we nonetheless find different behavior for the predictive power and process complexity. To quantify any quantum advantage, we adopt quantum discord \cite{Ollivier01} as a measure of quantum correlations. 
\alg{We will see that coherences in the environment of the predictive system can improve predictive efficiency}, to a degree quantified by a negative (quantum) part of the lost work. This result assigns an operational meaning to quantum discord: it is the thermodynamic inefficiency of the most energetically efficient classical approximation of a quantum memory.

Central to our results is the extension of Landauer's principle to quantum systems recently made by Rio \etal\ \cite{Rio11}. Landauer's principle states that the erasure of information from a memory must necessarily lead to the generation of heat, at minimum in proportion to its entropy \cite{Landauer61}. The converse of this statement, which will be of use to us, is that any memory not in a state of maximum entropy can \emph{in principle} be used to perform work. In extending this to quantum systems, one must take into account any ``quantum side information'' held about the system in its environment \cite{Rio11}.

Let $S$ be a quantum system, and denote by $X$ some part of the surrounding environment that is correlated to $S$; we assume that $S$ is finite and consists of $N$ qubits. The setup might represent some physical system responding to a changing environment whose state is encoded in $X$. From a thermodynamic point of view, we will view information held in $S$ as a potential source of work.

Consider now that $X$ undergoes an evolution, $\rho_X \to \rho_X' = \mathcal{E}(\rho_X)$, with $\mathcal{E}$ a quantum channel. Meanwhile, information about $X$ is held by $S$, and some of this information is useful for making predictions about the evolution. The two systems are represented by quantum state $\rho_{SX}$, and the evolution of $X$ can be written as a local map: $\rho_{SX} \to \rho_{SX}' = I_S\otimes\mathcal{E}(\rho_{SX})$, where $\tr_S\rho_{SX} = \rho_X$, $\tr_S\rho_{SX}' = \rho_X'$, and $I_S$ is the identity on $S$. Different $\rho_{SX}$, constrained by $\tr_S\rho_{SX} = \rho_X$, represent different models encoding information about $X$. If we are interested in making predictions about the future state, a model might contain redundant information at the level of accuracy required. We can quantify information about the past state of $X$, which we refer to as the \emph{memory}, by the mutual information $I(S:X) := H(\rho_S) - H(S|X)$, where $H(\rho) = -\tr[\rho\log_2\rho]$ is the von-Neumann entropy measured in bits, $H(S|X) = H(\rho_{SX}) - H(\rho_X)$ is the conditional quantum entropy of $S$ given $X$, and $\rho_S$ denotes the reduced state of $S$. Similarly, the \emph{predictive power}, the information we have about the future state, can be quantified by the mutual information $I(S:X') := H(\rho_S) - H(S|X')$, where $H(S|X') = H(\rho_{SX}')-H(\rho_X')$. If two models achieve the same predictive power, the one containing the least information about the past displays a more efficient use of information.

Recently, Still \emph{et al.} \cite{Still12} proposed that the difference $I(S:X) - I(S:X')$ serves as a natural and meaningful measure of predictive inefficiency. We may view it as non-predictive information. Classically, the optimization of this quantity can be seen as a special case of the information bottleneck method, where $I(S:X)$ is minimized at fixed $I(S:X')$ \cite{Tishby99,Still10}. It was also shown in \cite{Still12}, that for classical systems, this measure equals the work lost when one considers extractable work from $S$ at a certain inverse temperature $\beta = 1/k_B T$. We show that the same identity holds for quantum systems.

As the state of $X$ is changed, work is performed on the total system $\rho_{SX}$. We then ask: does the change in $X$ change the work recoverable from $S$? Assuming all states are energetically equivalent---\emph{i.e.}, a fully degenerate Hamiltonian---it was shown in \cite{Rio11} that one can extract an amount of work $W_\text{ext}[S|X] = \beta^{-1}[N - H(S|X)] \ln 2$ from $S$, at inverse temperature $\beta$, with the state of $X$ kept intact \cite{note1}. This holds in the asymptotic limit where work extraction is considered on a large number of copies of the system. We adopt this limit throughout this paper (see \cite{Egloff12} for a discussion). We then define a thermodynamic inefficiency for $S$, at inverse temperature $\beta$, as the difference of $W_\text{ext}[S|X]$ and $W_\text{ext}[S|X']$, i.e., the negative change in work that can be extracted from $S$ keeping the surroundings intact:
\begin{equation}\label{eq:W_diss1}
  \beta W_\text{lost}[\rho_{SX}\to\rho_{SX}']:= [H(S|X') - H(S|X)]\ln 2.
\end{equation}
It follows that
\begin{equation}\label{eq:W_diss2}
  \beta W_\text{lost}[\rho_{SX}\to\rho_{SX}']=[I(S:X) - I(S:X')]\ln 2,
\end{equation}
which shows that, just as in the classical case, predictive inefficiency equals thermodynamic inefficiency. Note that $W_\text{lost}$ is positive, as mutual information cannot increase over a local map on $X$ (the data processing inequality, see for example \cite{NielsenChuang04}) \cite{note3}. This result, which is analogous to the main result from \cite{Still12}, follows directly from the expression for extractable work from \cite{Rio11}, but note that the results in \cite{Rio11} are derived from rather different principles than those used in \cite{Still12}. In particular, central to the latter, is the definition of a non-equilibrium free-energy. One could define an analogous quantity for quantum systems, \emph{i.e.}, $\beta F_{\rm neq}(S|X) = \beta \braket{H_S} - H(S|X)$, where $H_S$ is a Hamiltonian, but such a definition clearly requires an operational interpretation. In general, it is not clear how free-energy should be defined for quantum systems with ``side information''. It is therefore not obvious that an expression for the lost work, identical to the classical expression, should exist for quantum systems as well \cite{note4}.

Equation~(\ref{eq:W_diss1}) involves quantum correlations. It is thus not surprising that quantum effects enter into the effective use of information and thermodynamic efficiency as we define them. Quantum discord \cite{Ollivier01} makes this explicit. It measures how shared quantum information deviates from classical mutual information. For a definition we first introduce a ``semiclassical'' conditional entropy---a measure of the uncertainty left about $S$ when a rank one projective measurement is performed on $X$ (or vice versa). Denoting the measurement by $\{P_k\}$, with $k$ an outcome, $\rho_k = I_S \otimes P_k \rho_{SX} I_S \otimes P_k/p_k$ is the system state conditioned on $k$, with $p_k = \tr[I_S \otimes P_K \rho_{SX}]$ the outcome probability. Now define $H(S|X=k):= H(\rho_k)$ as the conditional entropy of $S$ given outcome $k$ measured on $X$, and introduce $H(S|X^C) := \min \sum_k p_k H(S|X=k)$ as a conditional entropy, where the minimum is taken over all rank one projective measurements. This quantity may be seen as an alternative to the usual quantum conditional entropy, $H(S|X)$, generalized from the conditional Shannon entropy for two classical random variables \cite{Ollivier01}. Quantum discord is defined as the (positive) difference $\delta(S|X) := H(S|X^C) - H(S|X)$.

We use $\delta(S|X)$ to quantify how quantum correlations enter into the thermodynamic efficiency of our model, as given by \eq{eq:W_diss1}; conversely, through \eq{eq:W_diss1}, we assign a \alg{new}, thermodynamic, interpretation to quantum discord. Note, first, that approximating $X$ as classical can be viewed as sending it down a decoherence channel, $\rho_{SX} \to \rho_{SX}^{\text{decoh}} = \sum_k I_S \otimes P_k \rho_{SX} I_S \otimes P_k$. Consider now the channel that minimizes the work lost according to \eq{eq:W_diss1}. From the definition of quantum discord,
\begin{align}\label{eq:discord_interp}
  \min \beta W_\text{lost}[\rho_{SX} \to \rho_{SX}^{\text{decoh}}] = \delta(S|X) \ln 2,
\end{align}
where the minimum is taken over all decoherence channels acting locally on $X$. 
In other words, quantum discord gives the energetic inefficiency, measured as lost work potential, of the most energetically efficient classical approximation of $X$, when viewing the information held by $S$ as a source of work. \alg{We can think of $X$ as a memory containing information about $S$ (just as $S$ can be thought of as a memory of $X$), and discord is therefore the minimum lost work when approximating a quantum memory as classical.} It is also then equal to the non-predictive information in this decoherence process. See \cite{Zurek03,Modi10,Cavalcanti11,Madhok11,Gu12} for previous interpretations of quantum discord.

Considering mutual information written as $I(S:X) = H(\rho_S) - H(S|X)$, we introduce as a natural measure of classical correlations, $I^C(S|X) := H(\rho_S) - H(S|X^C)$ \cite{Henderson01}. We then divide the total correlations into a ``classical'' and a ``quantum'' part: $I(S:X) = I^C(S|X) + \delta(S|X) = I^C(X|S) + \delta(X|S)$, where we define $I^C(X|S)$ and $\delta(X|S)$ with the projective measurement performed on $S$ instead of $X$. Of course, we can divide the predictive information, $I(S:X')$, into parts in exactly the same manner. Importantly, in general $\delta(S|X) \ne \delta(X|S)$, and either of these two quantities greater than zero signals non-classical correlations.

The lost work acquires classical and quantum parts from its dependence on $I(S:X)$ and $I(S:X')$ [\eq{eq:W_diss2}]:
\begin{eqnarray}\label{eq:W_diss3}
  W_{\text{lost}}[\rho_{SX}\to\rho_{SX}'] &=& W^C_{\text{lost}}(S|X) + W^Q_{\text{lost}}(S|X) \\
                                          &=&W^C_{\text{lost}}(X|S) + W^Q_{\text{lost}}(X|S), \nonumber
\end{eqnarray}
where 
\alg{$\beta W^{C}_{\text{lost}}(S|X) := [I^{C}(S|X) - I^{C}(S|X')]\ln 2$, }
\alg{$\beta W^{Q}_{\text{lost}}(S|X) := [\delta(S|X) - \delta(S|X')]\ln 2$, }
and an analogous expression applies when the projective measurement is performed on $S$.

In the quantum case one typically expects to have non-zero quantum discord both before and after the map on $X$ \cite{Ferraro10}. In other words, there is typically some quantum advantage in predictive power, $I(S:X')$, though, at the same time, a larger process complexity, $I(S:X)$, due to quantum correlations. There is an advantage in thermodynamic efficiency only when $W^Q_{\text{lost}}<0$. This is possible because discord can be created by a local map \cite{Streltsov11,Ciccarello12}, an important contrast with entanglement. Indeed, the map $I_S\otimes{\mathcal E}$, cannot increase entanglement \cite{Vedral97}. 
\alg{The map cannot generate quantum correlations of the kind measured by $\delta(X|S)$ either. $\delta(X|S)$ pertains to a situation where $S$ is considered the ``apparatus'', \emph{i.e.}, information about $X$ is inferred from a measurement on $S$---and the discord stems from $S$ not being in a classical state. As the map, $I_S\otimes\mathcal{E}$ acts locally on $X$, further discord cannot be created with $S$ considered the apparatus \cite{note2}. }
\alg{The same is not true for $\delta(S|X)$, where a measurement is considered on $X$, however, as orthogonal (distinguishable) quantum states for $X$ can be mapped onto non-orthogonal (indistinguishable) states by $\mathcal{E}_i$.}
Any quantum advantage must therefore be quantified by $W_{\text{lost}}^Q(S|X)$, and, in particular, no such advantage is possible for a system learning from a classical environment; although there might still be an advantage to the predictive power, as measured by $\delta(X'|S)$.

\begin{figure}[t]
  \includegraphics[scale=0.9]{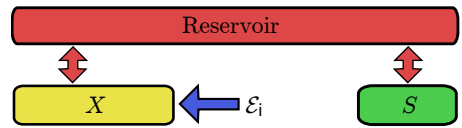}
\caption{\label{fig:evol} (Color online.) Schematic of the setup.
System $S$ gathers information about $X$ through a common reservoir interaction (map ${\mathcal R}_{SX}$), while $X$ is driven from equilibrium by the maps $\mathcal{E}_i$. The evolution of $\rho_X(t)$ is not modified by the presence of $S$. The changing of $S$ due to the evolution of $X$ is interpreted as an implicit computational model.}
\end{figure}

We now apply these results to a situation where the state, $\rho_{SX}$, prior to the update on $X$ results from a process where $S$ and $X$,  initially uncorrelated, interact through a common one-way reservoir. Potentially, the interaction can entangle and correlate the systems, and it is interesting to consider whether a quantum advantage to predictive efficiency can arise as information flows from $X$ to $S$. This is a close quantum analog of the process considered in \cite{Still12}. In this scheme (see Fig.~\ref{fig:evol}), the state of $X$ is updated at discrete times $t_0$, $t_1$, $\ldots$, with $\rho_{X}(t_i) \to \rho_{X}'(t_i) = \mathcal{E}_i(\rho_{X}(t_i))$. In between these changes, information can be read from $X$ through an interaction with a reservoir. 
For this we introduce the map $\mathcal{R}_X$. The evolution of $X$ is then written formally as $\dots\to\rho_X(t_i) \to \rho_X'(t_i) = \mathcal{E}_i(\rho_X(t_i))  \to  \rho_X(t_{i+1}) = \mathcal{R}_X(\rho_X'(t_{i})) \to \dots$.

The quantum mechanical analogue of a ``passive learning'' scenario  \cite{Still12}, where $S$ may not change $X$ (contrast an ``interactive learning'' scenario \cite{Still09}), is then a situation where $S$ retrieves information about $\rho_{X}'$ by interaction with the same reservoir in such a way that the evolution of $X$ is unchanged. In summary, we consider an evolution for $S$ plus $X$:
\begin{subequations}
\begin{align}
  &\rho_{SX}(t_i) \to \rho_{SX}'(t_i) = I_S \otimes \mathcal{E}_i(\rho_{SX}(t_i)) \label{eq:X_update_simple}, \\
  \noalign{\vskip3pt}
  &\rho_{SX}^\prime(t_i) \to \rho_{SX}(t_{i+1}) = \mathcal{R}_{SX}(\rho_{SX}^\prime(t_i)) \label{eq:relax},
\end{align}
\end{subequations}
where $I_S$ denotes the identity on $S$ and $\tr_S[\mathcal{R}_{SX}(\rho(t))] = \mathcal{R}_X(\rho_X(t))$.

Note that Eqs.~(\ref{eq:X_update_simple}) and (\ref{eq:relax}) imply a separation of time scales: the update on $X$ happens much faster than any timescale associated with the reservoir interaction and response of $S$; the state of $S$ is not changed during the update---$\rho_S(t_i) = \tr_X[\rho_{SX}(t_i)] = \tr_X[\rho_{SX}'(t_i)]$. We may then, as in the classical case, view \eq{eq:X_update_simple} as a (fast) ``work step'' and \eq{eq:relax} as a (slow) ``relaxation step''.

We note two further considerations. First, there is a departure from the assumption of the classical setup that the state of $X$ not change during the relaxation  \cite{Still12}. In the quantum case the state of $X$ can change. As a result, we do not consider the thermodynamic operation of the process as a whole, over a sequence of many time steps, but limit ourselves to the thermodynamic and predictive inefficiency at a single update. Our interest lies with the predictive capabilities of $S$ at time $t_i$, when the model encoding the predictive information, $\rho_{SX}(t_i)$, results from
the evolution up to time $t_i$.

Second, it is important to note that we consider a part of $W_\text{lost}(t_i) := W_\text{lost}[\rho_{SX}(t_i)\to\rho_{SX}'(t_i)]$ that arises from genuine quantum correlations, present before and after each update. This should not be confused with a situation where $X$ (or $S$) evolves according to a classical stochastic process. That would entail a state $\rho_{SX}(t) = \sum_k p_k(t) \rho_k(t) \otimes P_k$ with $p_k(t)$ that stochastic process, where $\rho_k(t)$ are density matrices for $S$, and $P_k$ are rank one projectors for $X$. Importantly, the basis $\{P_k\}$ is fixed for all times in this case, whereas the basis achieving the minimum in the definition of discord might be different at different times.

Turning now to an illustration, we consider an example where quantum correlations contribute positively to the predictive efficiency. With both $S$ and $X$ taken to be qubits, we consider a local map on $X$ shown in \cite{Ciccarello12} to produce quantum correlations: ${\mathcal E}_i(\rho)=K_0\rho K_0^\dagger+K_1\rho K_1^\dagger$, with Krauss operators $K_0 = \ket{0}\bra{0}_X + \sqrt{1-p}\ket{1}\bra{1}_X$ and $K_1 = \sqrt{p}\ket{0}\bra{1}_X$; and we set $p=0.7$. \alg{These maps model the stochastic evolution of the environment, driving the system out of equilibrium.} We adopt the formalism of cascaded quantum systems \cite{Carmichael93_2} to describe the response of $S$. Under this scheme, qubit $X$ couples to a zero temperature one-way reservoir $R$ at position $z=0$, with interaction Hamiltonian $H_{XR} = i \sqrt{2\kappa}(\sigma_x^X\mathcal{E}\dagg(0) - {\rm H.c.})$, while $S$ couples to $R$ at $z=l$, with interaction  $H_{SR} = i \sqrt{2\kappa}(\sigma_-^S\mathcal{E}\dagg(l) - {\rm H.c.})$; $2\kappa$ is the interaction rate for each qubit, $\sigma_x^X$ and $\sigma_-^S$ are Pauli operators, and $\mathcal{E}(z)$ is the (photon) annihilation operator at $z$ for the reservoir mediating the one-way coupling of $X$ to $S$. The self-Hamiltonians of the qubits are set to zero, $H_S = H_X = 0$. We let $l\to 0$, for simplicity, and describe the system response through the master equation \cite{Carmichael93_2} ($\hbar = 1$)
\begin{eqnarray}\label{eq:cascaded}
\dot{\rho}&=& -i[H,\rho] +\mathcal{D}[C]\rho_{SX},
\end{eqnarray}
where $H = i\kappa\sigma_x^X(\sigma_-^S - \sigma_+^S)$, and $C = \sqrt{2\kappa}(\sigma_x^X + \sigma_-^S)$; the superoperator $\mathcal{D}$ is defined through $\mathcal{D}[O]\rho = O\rho O\dagg - O\dagg O\rho/2 - \rho O\dagg O/2$.

\begin{figure}
\includegraphics[width=3.2in]{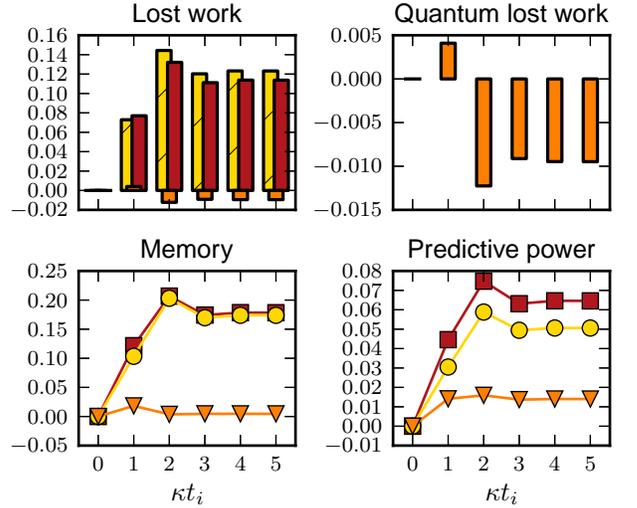}
\caption{\label{fig:Wdiss} (Color online.) Top left: Lost work in the implicit computation at the discrete set of times $\kappa t_i$; $\beta W_\text{lost}(t_i)/\ln 2$ (dark red [dark gray] bars, positive values), $\beta W_\text{lost}^C(t_i)/\ln 2$  (yellow [light gray] dashed bars, positive values), $\beta W_\text{lost}^Q(t_i)/\ln 2$ (orange [light gray] bars, smallest magnitudes). Top right: The quantum part, $\beta W_\text{lost}^Q(t_i)/\ln 2$ only. Bottom left: Memory at times $\kappa t_i$: $I(S:X;t_i)$ (dark red squares), $I^C(S:X;t_i)$ (yellow circles), $\delta(S:X;t_i)$ (orange triangles). Bottom right: Predictive power at times $\kappa t_i$: $I(S:X';t_i)$ (dark red squares), $I^C(S:X';t_i)$ (yellow circles), $\delta(S:X';t_i)$ (orange triangles).}
\end{figure}

Starting from an uncorrelated initial state, $\rho_{SX}(t_0) = \ket{0}\bra{0}_S\otimes\left(\ket{+}\bra{+}_X + \ket{-}\bra{-}_X\right)/2$,  $\ket{\pm} = (\ket{0}\pm\ket{1})/\sqrt2$, the evolution generates correlations in the system. We choose $\kappa(t_{i+1}-t_i) = 1$, for all $i$, and calculate numerically $W_\text{lost}(t_i)$, $W^C_\text{lost}(t_i)$, and $W^Q_\text{lost}(t_i)$, considering projective measurements on $X$, \emph{i.e.}, $W^{C,Q}_\text{lost}(t_i) := W^{C,Q}_\text{lost}(S|X)$, under $\rho_{SX}(t_i) \to \rho_{SX}'(t_i)$. The results appears in \fig{fig:Wdiss}. \alg{We also include in the figure the memory $I(S:X;t_i)$ 
and the predictive power $I(S:X';t_i)$.}

As expected the total and classically lost work are positive, while the quantum part is negative for all $t_i>t_1$. The efficiency is zero in the uncorrelated initial state, beyond which the relaxation process, \eq{eq:cascaded}, creates correlations and allow the model to be predictive; it may be thought of as the ``learning'' part of the implicit computation. \alg{More specifically, the relaxation process given in \eq{eq:cascaded} maps the initial state onto an ``X state'', \emph{i.e.}, a state of the form
\begin{align}
\rho_ = \left( \begin{array}{cccc}
\rho_{11} & 0 & 0 & \rho_{14} \\
0 & \rho_{22} & \rho_{23} & 0 \\
0 & \rho_{32} & \rho_{33} & 0 \\
\rho_{41} & 0 & 0 & \rho_{44} \end{array}\right).
\end{align}
Note, in particular, that the steady state of the relaxation is given by $\rho_{11} = \rho_{33} = 2/9$, $\rho_{22} = \rho_{44} = 5/18$, $\rho_{14} = \rho_{23} = \rho_{32} = \rho_{41} = -1/9$. The map $\mathcal{E}_i(\rho)$ then takes this state to
\begin{align*}
\rho \to \left( \begin{array}{cccc}
(1-p)\rho_{11} & 0 & 0 & \sqrt{1-p}\rho_{14} \\
0 & (1-p)\rho_{22} & \sqrt{1-p}\rho_{23} & 0 \\
0 & \sqrt{1-p}\rho_{32} & \rho_{33}+p\rho_{11} & 0 \\
\sqrt{1-p}\rho_{41} & 0 & 0 & \rho_{44}+p\rho_{22} \end{array}\right);
\end{align*}
thus the evolution is restricted to the subspace of X-states. The map $\mathcal{E}_i$ maps to a state with higher discord when applied to the steady state of the relaxation process as given above. The relaxation time ($\kappa(t_{i+1}-t_i)=1$) is not quite long enough for the system to reach the steady state before each update during the first few steps, but it eventually does so, and the disspated work thus reaches a steady state value. Also, we mention that we find the basis minimizing the discord to be the computational basis, $\{\ket{0}_X,\ket{1}_X\}$, for all $t_i$.}

We have shown that as a quantum system $X$ changes state, the non-predictive quantum information held by a correlated system $S$ equals the lost work potential from $S$. In particular, if $X$ changes to a classical state which minimizes the lost work, this lost work is given by the quantum discord present before the change. More generally, quantum discord quantifies the contribution to the lost work coming from quantum correlations. We have demonstrated that such correlations can contribute positively to the thermodynamic operation.

\alg{We would also like to point to some open questions in light of the present discussion. The ``classicality'' of what we call the classically lost work, $W_\text{lost}^C$, is open to question, since classicality would entail $X$ being described by a classical stochastic process, as already pointed out. It would therefore be interesting to compare predictive quantum processes to optimal \cite{Still10} classical counterparts; and in a similar vein, consider \emph{optimal} predictive quantum systems. We leave these considerations to a future work.}

\begin{acknowledgments}
The author would like to thank Howard Carmichael, Scott Parkins, Lidia del Rio, Bo-Sture K. Skagerstam and Susanna Still for helpful discussions.
\end{acknowledgments}


\begin{thebibliography}{24}
\bibitem{note1} Being ``kept intact'' means that, assuming $S$, $X$, and the rest of the enivronment, say $R$, make up a total pure state, then the reduced state of $X+R$ is not affected by the work extraction.
\bibitem{note3}Interestingly, we note that $W_\text{lost}$ as defined here can be negative due to entanglement generation if one allows global maps; for example, if $\rho_{SX}$ is pure, $H(S|X)$ is negative if and only if the state is entangled; it follows that if the quantum channel maps a pure unentangled state onto a pure entangled state, $W_\text{lost} <0$, and net work is gained.
\bibitem{note4} If formulating thermodynamics as a resource theory \cite{Brandao11}, it is natural in the present context to consider the ``free state'' to be $\sigma_\beta \otimes \rho_X$ where $\sigma_\beta = \exp(-\beta H_S)/\tr[\exp(-\beta H_S)]$ is the Gibbs state at inverse temperature $\beta$ for $S$, and $\rho_X$ is the fixed state for $X$. In this case the quantity $\beta F_\text{neq}(S|X) = \beta\braket{H_S} - H(S|X)$ does indeed play the same role as the usual free energy, $F(S) = \beta\braket{H_S} - H(S)$ (when no side-information is to be preserved), in that it measures the relative entropy to the ``free state'', $H(\rho_{SX}||\sigma_\beta\otimes\rho_X)$. See \cite{Brandao11} for details.
\bibitem{note2}Write \cite{Modi10}  $\delta(X|S) = H(\rho_{SX}||\chi) - H(\sum p_{k} \ket{k}\bra{k}) + H(\rho_S)$,  where $H(\rho_{SX}||\chi)$ is a relative entropy and $\chi = \sum p_{k} \ket{k}\bra{k} \otimes \rho_k^X$ is the state that minimizes the discord. Define $\chi' = I_S\otimes\mathcal{E}(\chi)= \sum p_{k} \ket{k}\bra{k} \otimes \mathcal{E}(\rho_k^{X})$, and $\delta(X'|S)_{\chi'} := H(\rho_{SX}'||\chi') - H(\sum p_{k} \ket{k}\bra{k}) + H(\rho_S)$. Note that relative entropy is monotonic under a completely positive trace preserving map: $H(\mathcal{E}(\rho)||\mathcal{E}(\sigma)) \le H(\rho||\sigma)$ \cite{NielsenChuang04}. It follows that $\delta(X'|S)_{\chi'} \le \delta(X|S)$. From this the claimed result is proved since $\delta(X'|S) \le \delta(X'|S)_{\chi'}$ by construction.
\end{thebibliography}

\end{document}